\documentclass[twocolumn,letter]{jpsj3}

\usepackage{txfonts}
\usepackage{bm}
\usepackage{color}%

\def\up{\uparrow}
\def\down{\downarrow }
\def\Vec#1{\bm{#1}}

\begin{document}


\title{Field-angle-dependent Low-energy Excitations around a Vortex in the Superconducting Topological Insulator Cu$_{x}$Bi$_{2}$Se$_{3}$}

\author{Yuki \surname{Nagai}}
\inst{
\address{CCSE, Japan  Atomic Energy Agency, 5-1-5 Kashiwanoha, Kashiwa, Chiba 277-8587, Japan} 
}

\date{\today}
             
\abst{
We study the quasiparticle excitations around a single vortex in the superconducting topological insulator Cu$_{x}$Bi$_{2}$Se$_{3}$, focusing on a superconducting state with point nodes. 
Inspired by the recent Knight shift measurements, we propose two ways to detect the positions of point nodes, 
using an explicit  formula of the density of states with Kramer-Pesch 
approximation in the quasiclassical treatment. 
The zero-energy local density of states around a vortex parallel to the $c$-axis has a twofold shape and 
splits along the nodal direction with increasing energy; 
these behaviors 
can be detected by the scanning tunneling microscopy. 
 An angular dependence of the density of states with a rotating magnetic field on the $a$-$b$ plane has deep minima when the magnetic field is parallel to the 
 directions of point nodes, which can be detected by angular-resolved heat capacity and thermal conductivity measurements. 
 All the theoretical predictions are detectable via standard experimental techniques in magnetic fields. 
}
%
\maketitle
 The discovery of topological insulators and superconductors 
 attracts a great deal of attention in condensed matter physics.
 A nontrivial 
 topological feature of the bulk state shows up in the edge boundaries  with closing of 
 insulating or superconducting gaps.\cite{Bernevig15122006,PhysRevLett.105.266401,PhysRevB.76.045302,PhysRevLett.98.106803,RevModPhys.82.3045,PhysRevLett.95.146802,Konig02112007,PhysRevLett.105.146801,PhysRevB.75.121306,PhysRevB.81.041309,PhysRevLett.105.136802} 
The Majorana fermion (a gapless zero-energy quasiparticle) in topological superconductors is a curious particle whose annihilation and creation operators are identical. 
A quest for the Majorana fermions is now an exciting issue. 

A typical topological insulator Bi$_{2}$Se$_{3}$ characterized by a $Z_{2}$ topological invariant shows superconductivity 
with Cu intercalating\cite{PhysRevLett.104.057001,Wray2010,MKR_L11,PhysRevLett.105.097001}. 
The copper-doped bismuth-selenium compounds are candidates for bulk topological superconductors. 
Although their properties are studied actively,\cite{PhysRevLett.107.217001,PhysRevB.86.064517, PhysRevLett.110.117001} 
identifying the gap-function type is an unsettled issue.  
Several groups observed zero-bias conductance peaks (ZBCPs) by point contact spectroscopy\cite{PhysRevLett.107.217001,PhysRevB.86.064517}. 
The measured ZBCPs can be evidence of topological superconductivity, since 
the topologically protected gapless Majorana fermions form at the edges. 
However, Levy {\it et al.} reported scanning tunneling spectroscopy (STS) measurements in  Cu$_{0.2}$Bi$_{2}$Se$_{3}$. 
The tunneling spectrum shows that the density of states (DOS) at the Fermi level 
is fully gapped, 
without any in-gap states; a fully-gapped non-trivial superconductivity occurs in Cu$_{0.2}$Bi$_{2}$Se$_{3}$.  
Thus, clear evidence of the topological superconductivity is now in great demand. 

The recent Knight-shift measurements\cite{Matano} show the presence of in-plane anisotropy in Cu$_{x}$Bi$_{2}$Se$_{3}$; 
this anisotropy can be related to the nodal characters of the superconducting state on $a$-$b$ plane\cite{Hashimoto}, since 
the electronic structure in normal states is isotropic. 
A theoretical model\cite{PhysRevLett.105.097001,NagaiMajo} of this superconductor predicts that 
within on-site pairing interaction  there are four different pairing symmetries, classified by the $D_{3d}$ point-group representations ($A_{1g}$, $A_{1u}$, $A_{2u}$, and $E_{u}$). 
The odd-parity $A_{1u}$ state (so-called $\Delta_{2}$) is fully gapped, whereas the odd-parity $A_{2u}$ and $E_{u}$ states ($\Delta_{3}$ and $\Delta_{4}$, respectively) have 
the point nodes. 
Thus, the anisotropic Knight shift suggests that Cu$_{x}$Bi$_{2}$Se$_{3}$ is a topological superconductor with breaking the rotational isotropy. 

An anisotropy feature originating from nodes in a superconducting state is detectable in Cu$_{x}$Bi$_{2}$Se$_{3}$ by various ways. 
The angle dependence of the thermal conductivity in the basal $a$-$b$ plane predicts\cite{NagaiThermal}
distinct strong anisotropy in the $E_{u}$ representation, without magnetic fields.  
Detecting quasiparticle excitations around a vortex is also useful for determining point nodes.  
A pattern of the local density of states (LDOS) around a vortex, which can be detected by scanning tunneling microscopy/spectroscopy(STM/STS), 
 have strong anisotropy in unconventional superconductors, since the spectrum due to the Andreev bound states around a vortex is 
 sensitive to the anisotropy of the pair potential\cite{Hess,Nishimori,Gygi,Hayashi,NagaiJPSJ,NagaiJPSJCe}. 
Furthermore, thermal transport measurements such as  heat capacity and 
thermal conductivity measurements with a rotating magnetic field are probes for node positions in unconventional superconductors\cite{Sakakibara,Matsuda}.

Let us here discuss what is an efficient and intuitive way to seek the unconventional properties of the topological superconductor.
A clue is the quasiclassical approach of superconductivity\cite{NagaiQuasi}. 
The author showed that 
the odd-prity superconductivity in topological superconductors turns to the spin-triplet one in terms of the quasiclassical treatment. 
The original massive Dirac Bogoliubov-de Gennes (BdG) Hamiltonian derived from a tight-binding model represented by 
an $8 \times 8$ matrix is reduced to a $4 \times 4$ matrix. 
This indicates that low-energy nontrivial quasiparticle excitations in the topological superconductors 
can be analyzed using established theoretical techniques of the spin-triplet superconductors. 
With the use of the Riccati parametrization, equations of motion of quasiclassical Green's function represented by a $4 \times 4$ matrix 
becomes two Riccati-type differential equations represented by $2 \times 2$ matrices.  
In addition, if the effective gap function $\Delta_{\rm eff}$ is unitary ($\Delta_{\rm eff}^{\dagger} \Delta_{\rm eff} \propto 1_{2 \times 2}$),  
the quasiclassical Green's function is obtained by solving two Riccati-type differential equations with $1 \times 1$ {\it scalar} coefficients. 
We use Kramer-Pesch approximation (KPA) for these scalar Riccati equations to analyze 
the field-angle-dependent low-energy excitations around a vortex, which has been used in various kinds of unconventional 
superconductors\cite{NagaiJPSJCe,NagaiYNi,NagaiPRL}.

In this paper, we 
study  the quasicparticle excitations around a vortex  in 
the three-dimensional superconducting topological insulator Cu$_{x}$Bi$_{2}$Se$_{3}$, focusing on a superconducting state with point nodes.
Using the quasiclassical treatment, we derive Riccati-type first-order differential equations with scalar coefficients 
related to the original massive Dirac Hamiltonian.
The KPA to study the Andreev bound states
leads to an explicit formula of the LDOS around a single vortex. 
We propose two ways to detect the in-plane anisotropy of the topological superconductivity. 
In the magnetic field parallel to the $c$-axis, 
a twofold shape of the LDOS pattern around a vortex at zero energy splits along the nodal direction with increasing energy; this is 
detectable in the standard STM measurements.
In the magnetic field perpendicular to the $c$-axis, an angular dependence of the DOS on a rotating magnetic field on the $a$-$b$ plane 
has deep minima when the magnetic field is parallel to the directions of point nodes; 
the angular-resolved heat-capacity and thermal-conductivity measurements may show this behavior.

 We study a model of a topological superconductor in a three-dimensional system, with mean-field approximation. 
As for the normal parts,  the effective Hamiltonian with a strong spin-orbit coupling around the $\Gamma$ point 
is equivalent to that of the massive Dirac Hamiltonian with the negative Wilson mass term. 
  The Bogoliubov-de Gennes (BdG) mean-field Hamiltonian of the superconductivity is 
  $H = \int d\Vec{r} \Vec{\psi}^{\dagger}(\Vec{r}) {\cal H}(\Vec{r}) \Vec{\psi}(\Vec{r})$, with 
  the 8-component column vector $\Vec{\psi}(\Vec{r})^{T} = (c_{\Vec{r},1,\up},c_{\Vec{r},1,\down},
  c_{\Vec{r},2,\up},c_{\Vec{r},2,\down},c^{\dagger}_{\Vec{r},1,\up},c_{\Vec{r},1,\down}^{\dagger},
  c^{\dagger}_{\Vec{r},2,\up},c_{\Vec{r},2,\down}^{\dagger}
  )$. 
The corresponding BdG equations with $8 \times 8$ matrix eigenequations are\cite{NagaiThermal,NagaiQuasi,Hao,Yamakage,Mizushima}  
\begin{align}
\left(\begin{array}{cc}
h_{0}(\Vec{r}) & \Delta_{\rm pair}(\Vec{r})  \\
\Delta_{\rm pair}^{\dagger}(\Vec{r})  & - h_{0}^{\ast}(\Vec{r})
\end{array}\right)
\left(\begin{array}{c}
\Vec{u}(\Vec{r}) \\
\Vec{v}(\Vec{r}) 
\end{array}\right)
&= E \left(\begin{array}{c}
\Vec{u}(\Vec{r}) \\
\Vec{v}(\Vec{r}) 
\end{array}\right),
\end{align}
with the normal-state Hamiltonian 
\begin{equation}
h_{0}(\Vec{r})
= 
- \mu + \gamma^{0}
[
M_{0}
- i \partial_{x} \gamma^{1}  
- i \partial_{y} \gamma^{2}  
- i \partial_{z} \gamma^{3}  
],
\end{equation}
and the $4 \times 4$ pairing potential matrix $\Delta_{\rm pair}$. 
The $4 \times 4$ matrices $\gamma^{\mu}$ ($\mu = 0,1,2,3$) are the Gamma matrices in the 
Dirac basis. 
Using the Pauli matrices $\sigma^{i}$ and $s^{i}$ in orbital and spin spaces, 
we have $\gamma^{0} = \sigma^{3} \otimes s^{0}$ and $\gamma^{i} = i \sigma^{2} \otimes s^{i}$, 
 with a unit $2 \times 2$ matrix $s^{0}$.   
Within on-site interaction, the pair potential  $\Delta_{\rm pair}$ must fulfill the relation $\Delta_{\rm pair}^{T} = 
- \Delta_{\rm pair}$ owing to the fermionic property. 
We have six possible gap functions classified by a Lorentz-transformation property\cite{NagaiThermal}; 
they are classified into a pseudo-scalar, a scalar, and a polar vector (four-vector). 
A direction of point nodes is parallel to that of polar vectors, since the rotational symmetry is preserved around 
this direction (See, Appendix B in Ref.~\citen{NagaiThermal}). 
Thus, we propose a pair potential with point nodes in $\theta_{\rm N} $-direction on $a$-$b$ plane, 
\begin{align}
\Delta_{\rm pair}(\Vec{r}) &= i  (\cos \theta_{\rm N}  \gamma^{1} + \sin \theta_{\rm N}  \gamma^{2}) \gamma^{5} \gamma^{2} \gamma^{0}  f(\Vec{r}),\\
&=f(\Vec{r})  \left(\begin{array}{cccc}
0 & 0 & e^{i \theta_{\rm N} } & 0 \\
0 & 0 & 0 & -e^{-i \theta_{\rm N} } \\
- e^{i \theta_{\rm N} } & 0 & 0 & 0 \\
0 & e^{-i \theta_{\rm N} } & 0 & 0
\end{array}\right). \label{eq:deltapair}
\end{align}
Diagonalizing the BdG Hamiltonian in the uniform system, we find that the point-nodes are located at 
\begin{align}
\Vec{k}_{\rm node}^{\pm} = \pm \sqrt{\mu^{2} - M_{0}^{2} + |\Delta|^{2}}(\cos \theta_{\rm N} ,\sin \theta_{\rm N} ,0).
\end{align} 
The momentum-independent gap function $\Delta_{\rm pair}(\Vec{r})$ makes anisotropic energy spectrum, 
although the Fermi surface of the normal state is spherical in this model. 
In this paper, we focus on physics originating from the point nodes in the topological superconductor, from a 
phenomenological point of view, although the microscopic superconducting mechanism supporting the Knight-shift measurements 
is quite interesting.

Here, we use the quasiclassical approach, to reduce the $8 \times 8$ matrix to a $4 \times 4$ one\cite{NagaiQuasi}.
Using this approach, we show the fact that an odd-parity superconductivity in topological superconductors 
maps into a spin-triplet one. 
According to Ref.~\citen{NagaiQuasi}, 
the quasiclassical Andreev equations with $4 \times 4$ matrix eigenequations are 
\begin{align}
\left(\begin{array}{cc}
- i \Vec{v}_{\rm F} \cdot \Vec{\nabla} & \Delta_{\rm eff}(\Vec{r},\Vec{k}_{\rm F})  \\
\Delta_{\rm eff}^{\dagger}(\Vec{r},\Vec{k}_{\rm F})  &  i \Vec{v}_{\rm F} \cdot \Vec{\nabla} 
\end{array}\right)
\left(\begin{array}{c}
\Vec{f}(\Vec{r},\Vec{k}_{\rm F}) \\
\Vec{g}(\Vec{r},\Vec{k}_{\rm F}) 
\end{array}\right)
&= E \left(\begin{array}{c}
\Vec{f}(\Vec{r},\Vec{k}_{\rm F}) \\
\Vec{g}(\Vec{r},\Vec{k}_{\rm F}) 
\end{array}\right),
\end{align}
with the $2 \times 2$ effective pairing potential matrix $\Delta_{\rm eff}$. 
Considering the odd-parity gap functions, $\Delta_{\rm eff}(\Vec{r},\Vec{k}_{\rm F})$ is defined by $\Delta_{\rm eff}(\Vec{r},\Vec{k}_{\rm F}) = f(\Vec{r}) (\Vec{d}(\Vec{k}_{\rm F}) \cdot \Vec{s}) (i s^{2})$. 
The odd parity gap function with point nodes shown in Eq.~(\ref{eq:deltapair}) 
is represented by $\Vec{d}(\Vec{k}_{\rm F}) = (v_{{\rm F}z} \sin \theta_{\rm N} , -v_{{\rm F}z} \cos \theta_{\rm N} , v_{{\rm F}y} \cos \theta_{\rm N} - v_{{\rm F}x} \sin \theta_{\rm N} )$, with the Fermi velocity $\Vec{v}_{\rm F} \equiv \Vec{k}_{\rm F}/\sqrt{M_{0}^{2}+ |\Vec{k}_{\rm F}|^{2}}$. 
The corresponding quasiclassical Green's function is 
\begin{align}
\check{g} &= - \check{N} 
\left(\begin{array}{cc}
\hat{1} - \hat{a} \hat{b} & 2 i \hat{a} \\
- 2 i \hat{b} & -(\hat{1} + \hat{b} \hat{a})
\end{array}\right), \\
\check{N} &= \left(\begin{array}{cc}
(\hat{1} + \hat{a} \hat{b})^{-1} & 0 \\
0 & (\hat{1} + \hat{b} \hat{a})^{-1}
\end{array}\right),
\end{align}
where the $2 \times 2$ matrix coefficients $\hat{a}$ and $\hat{b}$ obey $2 \times 2$-matrix Riccati equations, 
\cite{NagaiJPSJCe,KatoJPSJ,Nagato,Higashitani,NagatoHigashitani,SchopohlMaki,Schopohl} 
\begin{align}
\Vec{v}_{\rm F} \cdot \Vec{\nabla} \hat{a} + 2 \omega_{n} \hat{a} + \hat{a} \Delta_{\rm eff}^{\dagger} \hat{a} - \Delta_{\rm eff} &= 0, \\
\Vec{v}_{\rm F} \cdot \Vec{\nabla} \hat{b} - 2 \omega_{n} \hat{b} - \hat{b} \Delta_{\rm eff} \hat{b} + \Delta_{\rm eff}^{\dagger} &= 0. 
\end{align}
Here, $\omega_{n}$ denotes the Fermion matsubara frequency. 
Considering clean superconductors in the type-II limit, we neglect the self-energy part of the Green's function and the vector potential. 
The following results do not change qualitatively, with taking impurity self-energies, as far as the point-nodes do not vanish in a dirty system.  
Assuming that the effective pairing potential matrix $\Delta_{\rm eff}$ is a unitary matrix ($\Delta_{\rm eff}^{\dagger} \Delta_{\rm eff} = |f(r)|^{2} |\Vec{d}(\Vec{k})|^{2} s^{0}$), we obtain scalar Riccati equations, 
\begin{align}
\Vec{v}_{\rm F} \cdot \Vec{\nabla} a' + 2 \omega_{n} a' + \Delta^{\ast} a^{'2}  - \Delta &= 0, \\
\Vec{v}_{\rm F} \cdot \Vec{\nabla} b' - 2 \omega_{n} b' - \Delta b^{'2} + \Delta^{\ast}&= 0,
\end{align}
where the scalars $a'$, $b'$ and $\Delta$ are defined by $\hat{a} \equiv a' (\Vec{d}(\Vec{k}_{\rm F}) \cdot \Vec{s}) (i s^{2})/|\Vec{d}(\Vec{k}_{\rm F})|$, $\hat{b} \equiv b' (\Vec{d}^{\ast}(\Vec{k}_{\rm F}) \cdot \Vec{s}^{\ast}) (i s^{2})/|\Vec{d}(\Vec{k}_{\rm F})|$ and 
$\Delta \equiv f(\Vec{r}) |\Vec{d}(\Vec{k}_{\rm F})|$, respectively. 
Since these equations contain $\Vec{\nabla}$ only through $\Vec{v}_{\rm F} \cdot \Vec{\nabla}$, 
the above equations reduces to a one-dimensional problem on a straight line, the direction of which 
is given by that of the Fermi velocity $\Vec{v}_{\rm F}$. 
Considering the cylindrical coordinate frame $(r,\alpha,z)$ with $\hat{z} \parallel \Vec{H}$ in the real space, 
the pair potential $\Delta$ does not depend on $z$ 
in the Riccati equations, and hence the Riccati equations can be rewritten as 
\begin{align}
v_{\rm F \perp}(\Vec{k}_{\rm F}) \frac{\partial a'}{\partial s} + 2 \omega_{n} a' + \Delta^{\ast}(s,y,\Vec{k}_{\rm F}) a^{'2}  - \Delta(s,y,\Vec{k}_{\rm F}) &= 0, \\
v_{\rm F \perp}(\Vec{k}_{\rm F}) \frac{\partial b'}{\partial s}- 2 \omega_{n} b' - \Delta(s,y,\Vec{k}_{\rm F}) b^{'2} + \Delta^{\ast}(s,y,\Vec{k}_{\rm F})&= 0,
\end{align}
where $v_{\rm F \perp}(\Vec{k}_{\rm F})$ is an amplitude of the vector $\Vec{v}_{\rm F \perp}$ perpendicular to the $z$ axis taken by projecting the 
Fermi velocity. 
Here, the argument $s(y)$ is along the direction parallel (perpendicular) to $\Vec{v}_{\rm F \perp}$.

Let us introduce the KPA as an efficient method to analyze the angle-resolved experiments. 
The zero-energy density of states around a vortex given by KPA is consistent with 
that of direct numerical calculations, as seen, {\it e.g.}, in Ref.~\citen{NagaiPRL}.
We expand  the Riccati equations up to first order with respect to energy and the imaginary part of the pair function\cite{Melnikov,NagaiOrganic}. 
For a single vortex, we adopt the spatial variation of the pair potential expressed by 
\begin{align}
f(\Vec{r}) = \Delta_{\infty} e^{i \alpha}  \frac{r}{\sqrt{r^{2} + \xi^{2}} },
\end{align}
with the cylindrical coordinate $\Vec{r} = (r,\alpha,z)$. 
The LDOS with the use of KPA is given as 
\begin{align}
n(\epsilon+ i \eta,s,y) &= \int \frac{d S_{\rm F}}{2 \pi^{3} |\Vec{v}_{\rm F}|} 
\frac{v_{\rm F \perp}}{ C(y,\Vec{k}_{\rm F})} 
\frac{
\eta e^{- u(s,\Vec{k}_{\rm F})}
}
{(\epsilon - E(y,\Vec{k}_{\rm F}))^{2} + \eta^{2}
},
\end{align}
where $\eta$ is the smearing factor, $d S_{\rm F}$ is the Fermi-surface area element, and 
\begin{align}
C(y,\Vec{k}_{\rm F}) &\equiv 2 \sqrt{y^{2} + \xi^{2}} K_{1} (r_{0}(y,\Vec{k}_{\rm F})), \\
E(y,\Vec{k}_{\rm F}) &= |\Vec{d}(\Vec{k}_{\rm F})| \Delta_{\infty} \frac{y}{\sqrt{y^{2} +\xi^{2}}} \frac{
K_{0}(r_{0}(y,\Vec{k}_{\rm F})) 
}
{
K_{1}(r_{0}(y,\Vec{k}_{\rm F})) 
}, \\
u(s,y,\Vec{k}_{\rm F}) &\equiv 
 \frac{
2 |\Vec{d}(\Vec{k}_{\rm F})| \Delta_{\infty}
}{
v_{\rm F \perp}
}
\sqrt{s^{2}+y^{2}+\xi^{2}}.
\end{align}
The quantity $r_{0}$ is defined by 
\begin{align}
r_{0}(y,\Vec{k}_{\rm F}) &= \frac{
2 |\Vec{d}(\Vec{k}_{\rm F})| \Delta_{\infty}
}{
v_{\rm F \perp}(\Vec{k}_{\rm F}) 
} \sqrt{y^{2} + \xi^{2}},
\end{align}
where the function $K_{n}(x)$ is the modified Bessel function of the second kind. 

\begin{figure}[t]
\begin{center}
\begin{tabular}{p{ \columnwidth}} 
\resizebox{ \columnwidth}{!}{\includegraphics{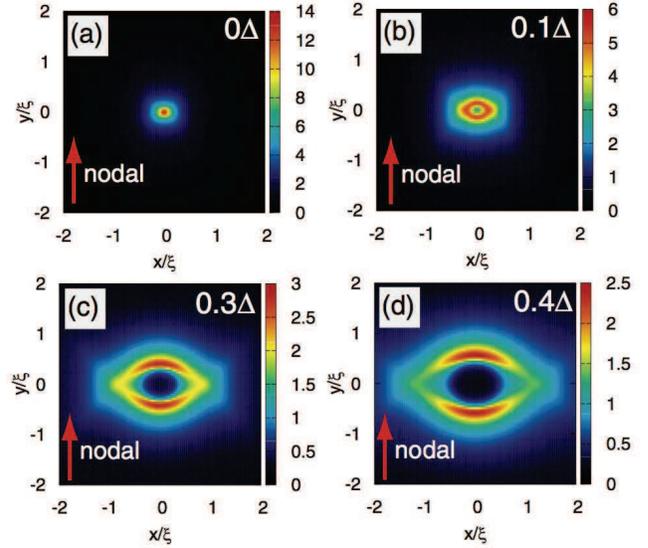}} 
\end{tabular}
\end{center}
\caption{\label{fig:fig1} 
(Color online) Energy dependence of the local density of states in a magnetic field parallel to $c$-axis. 
A direction of nodal quasiparticles is denoted as arrows in each panel. 
The peak position in Fig.1(c) and Fig.1(d) is located on the $y$ axis. 
} 
\end{figure}

Now, we study the LDOS pattern around a vortex in the magnetic field parallel to the $c$-axis. 
For simplicity, we adopt $y$-Polar-vector type gap function (so-called $\Delta_{4}$ in Ref.~\citen{PhysRevLett.107.217001}), where the point nodes are 
located on the $k_{y}$ axis ($\theta_{\rm N} = \pi/2$).  
We set the smearing factor $\eta = 0.05\Delta_{\infty}$  
Near zero energy, the bound states spread to the anti-nodal direction as shown in Fig.~\ref{fig:fig1}(a) and Fig.~\ref{fig:fig1}(b). 
In contrast, with increasing energy, the bound states spread to the nodal direction as shown in Fig.~\ref{fig:fig1}(c) and Fig.~\ref{fig:fig1}(d).
These results indicate that a twofold shape of the LDOS pattern at zero energy splits along the nodal direction with increasing energy. 
In Figs.\ref{fig:fig1}(a)-(b), 
the shape is elliptic spreading in the $x$-direction. On the other hand, in Figs.\ref{fig:fig1}(c)-(d), the peak position is located on the $y$ axis. 
The shape of the LDOS pattern was discussed in terms of the enveloping curve of the quasiparticle paths\cite{NagaiJPSJ}. 
The LDOS in $d$-wave superconductors spreads to the nodal direction at the zero energy, since there is the asymptote of the trajectory in the nodal direction\cite{NagaiJPSJ}. 
However, the trajectory in this system is a parabola. 
The zero-energy LDOS spreads to the anti-nodal direction, since the anti-nodal quasiparticle paths are dense near the vortex because of the sharp curve of trajectory. 
The split  of the LDOS pattern is evidence of unconventional superconductivity, 
since they have been observed in the superconductors with the anisotropic superconducting gap structure such as NbSe$_{2}$\cite{Hess,HayashiPRL} and YNi$_{2}$B$_{2}$C\cite{Nishimori,NagaiYNi}. 

Next, we examine the angular-resolved DOS in the magnetic field perpendicular to the $c$-axis. 
The DOS is calculated by 
\begin{align}
N(\epsilon+ i \eta) &= \langle 
n(\epsilon+ i \eta,\Vec{r}) 
\rangle_{\rm SP}.
\end{align}
Here, $\langle \cdots \rangle_{\rm SP} \equiv \int_{0}^{r_{a}} r dr \int_{0}^{2 \pi} d\alpha /(\pi r_{a}^{2})$ is the real-space 
average around a vortex where $r_{a}/\xi = \sqrt{H_{c2}/H}$ [$H_{c2} \equiv \Phi_{0}/(\pi \xi^{2})$ and $\Phi_{0} = \pi r_{a}^{2} H$]. 
We set the spatial cutoff length $r_{a} = 5 \xi$, which is comparable to the neighboring vortex distance as the magnetic field 
$H \sim H_{c2}/25$. 
\begin{figure}[t]
\begin{center}
\begin{tabular}{p{0.85 \columnwidth}} 
\resizebox{0.85 \columnwidth}{!}{\includegraphics{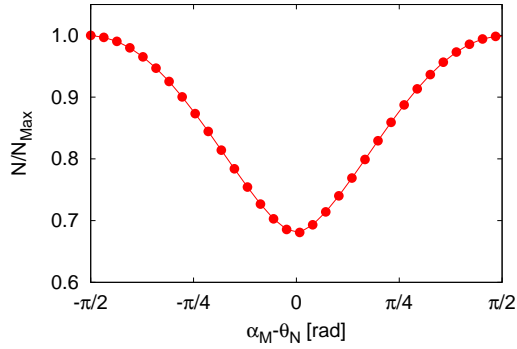}} 
\end{tabular}
\end{center}
\caption{\label{fig:fig2} 
(Color online) Angular dependence of the zero-energy density of states with a rotating magnetic field on the $a$-$b$ plane. 
The angle $\theta_{\rm N}$ denotes the direction of point nodes. 
} 
\end{figure}
The zero-energy DOS drastically decreases, when the magnetic field is applied in the parallel direction to the direction of the nodal 
quasiparticles ({\it i.e.}, $\alpha_{\rm M} = \theta_{\rm N}$), as shown in Fig.~\ref{fig:fig2}. 
Here, $\alpha_{\rm M}$ denotes the direction of magnetic fields. 
The order of this reduction is similar to 
the case of $d$-wave superconductivity with a cylindrical Fermi surface\cite{NagaiPRL}. 
%
In terms of the Doppler shift method, which has been frequently used to analyze experiments\cite{Sakakibara,Matsuda}, the quasiparticles around a vortex have the Doppler-shift energy 
$\delta E \propto \Vec{v}_{\rm F \perp} \cdot \Vec{v}_{s}$,  
where $\Vec{v}_{s}$ means the superfluid velocity around a vortex and $\Vec{v}_{s} \perp \Vec{H}$. 
The quasiparticles near the gap nodes mainly contribute to the DOS, since they are excited only by the Doppler-shift energy.  
In the magnetic field parallel to the gap-node direction, the nodal quasiparticles have no Doppler-shift energy, because $\Vec{v}_{\rm F \perp} = 0$. 
In terms of the KPA, the DOS originating from the bound states around a vortex reduces in the magnetic field parallel to the direction of the nodal quasiparticles, since the quasiparticles along the magnetic field are not bound. 
The angular-resolved DOS in Fig.~\ref{fig:fig2} can be observed by angular-resolved heat capacity and thermal conductivity measurements\cite{Sakakibara,Matsuda}, in topological superconductors with point-nodes.
In this paper, we do not take the contributions in the DOS originating from the surface bound states into account. 
Their contributions may be predominant in a small single crystal; indeed, the surface bound states lead to strong anisotropy in the thermal 
conductivity without a magnetic field, as seen in Ref.~\citen{NagaiThermal}.  
Our results indicate that the anisotropy from the point nodes can be detected by the angular-resolved DOS, even in the bulk,  with a rotating magnetic field on the $a$-$b$ plane. 

In conclusion, we proposed two ways to detect the position of point-nodes in the model of superconducting topological insulator Cu$_{x}$Bi$_{2}$Se$_{3}$ 
with the quasiclassical approach of superconductivity. 
We reduced $8 \times 8$ matrix eigenequations to two $1 \times 1$ scalar Riccati-type differential equations.
Furthermore, we applied the KPA, to obtain an explicit formula of the DOS. 
A twofold shape of the LDOS pattern around a vortex at zero energy splits along the nodal direction with increasing energy in the magnetic fields parallel to the $c$-axis. 
The DOS with a rotating magnetic field on the $a$-$b$ plane has deep minima in the magnetic field parallel to the directions of point nodes. 
All the theoretical predictions are detectable via standard experimental techniques, including the STM and the angler-resolved thermal transport measurements.

We thank Y. Ota, H. Nakamura and M. Machida for helpful discussions and comments. 
The calculations were performed using the supercomputing system PRIMERGY BX900 at the Japan Atomic 
Energy Agency. 
This study was supported by JSPS KAKENHI Grant Number 26800197.


\begin{thebibliography}{9}

\bibitem {Bernevig15122006}%
  B.~A.\ Bernevig, T.~L.\ Hughes, \ and\ S.-C.\ Zhang:\ Science\ \textbf
  {314}\ (2006) 1757.
\bibitem {PhysRevLett.105.266401}%
  Y.~L.\ Chen, Z.~K.\ Liu, J.~G.\ Analytis, J.-H.\ Chu, H.~J.\ Zhang, B.~H.\
  Yan, S.-K.\ Mo, R.~G.\ Moore, D.~H.\ Lu, I.~R.\ Fisher, S.~C.\ Zhang,
  Z.~Hussain, \ and\ Z.-X.\ Shen:\ Phys. Rev. Lett.\ \textbf {105} (2010) \ 266401.
\bibitem {PhysRevB.76.045302}%
  L.~Fu\ and\ C.~L.\ Kane:\ Phys. Rev. B\ \textbf {76}\ (2007) 045302.
\bibitem {PhysRevLett.98.106803}%
  L.~Fu, C.~L.\ Kane, \ and\ E.~J.\ Mele:\ Phys. Rev. Lett.\ \textbf {98} \ (2007)
  106803.
  \bibitem {RevModPhys.82.3045}%
  M.~Z.\ Hasan\ and\ C.~L.\ Kane: \ Rev. Mod. Phys.\ \textbf {82}\ (2010) 3045, and references therein.
\bibitem {PhysRevLett.95.146802}%
  C.~L.\ Kane\ and\ E.~J.\ Mele:\ Phys. Rev. Lett.\ \textbf {95}\ (2005) 146802.
\bibitem {Konig02112007}%
  M.~K{\"{o}}nig, S.~Wiedmann, C.~Br{\"{u}}ne, A.~Roth, H.~Buhmann, L.~W.\
  Molenkamp, X.-L.\ Qi, \ and\ S.-C.\ Zhang:\ Science\ \textbf {318}\ (2007) 766.
\bibitem {PhysRevLett.105.146801}%
  K.~Kuroda, M.~Ye, A.~Kimura, S.~V.\ Eremeev, E.~E.\ Krasovskii, E.~V.\
  Chulkov, Y.~Ueda, K.~Miyamoto, T.~Okuda, K.~Shimada, H.~Namatame, \ and\
  M.~Taniguchi: \ Phys. Rev. Lett.\ \textbf {105} (2010) \ 146801.
\bibitem {PhysRevB.75.121306}%
  J.~E.\ Moore\ and\ L.~Balents:\ Phys. Rev. B\ \textbf {75} (2007)\ 121306.
\bibitem {PhysRevB.81.041309}%
  A.~Nishide, A.~A.\ Taskin, Y.~Takeichi, T.~Okuda, A.~Kakizaki, T.~Hirahara,
  K.~Nakatsuji, F.~Komori, Y.~Ando, \ and\ I.~Matsuda:\ Phys. Rev. B\ \textbf
  {81}\ (2010) 041309.
\bibitem {PhysRevLett.105.136802}%
  T.~Sato, K.~Segawa, H.~Guo, K.~Sugawara, S.~Souma, T.~Takahashi, \ and\
  Y.~Ando:\ Phys. Rev. Lett.\ \textbf {105} \ (2010) 136802.
  \bibitem {PhysRevLett.104.057001}%
  Y.~S.\ Hor, A.~J.\ Williams, J.~G.\ Checkelsky, P.~Roushan, J.~Seo, Q.~Xu,
  H.~W.\ Zandbergen, A.~Yazdani, N.~P.\ Ong, \ and\ R.~J.\ Cava,\ Phys. Rev.
  Lett.\ \textbf {104},\ 057001 (2010).
 \bibitem {Wray2010}%
  L.~A.\ Wray, S.-Y.\ Xu, Y.~Xia, Y.~S.\ Hor, D.~Qian, A.~V.\ Fedorov, H.~Lin,
  A.~Bansil, R.~J.\ Cava, \ and\ M.~Z.\ Hasan,\ Nat Phys\ \textbf {6},\ 855
  (2010).  
 \bibitem{MKR_L11}
M. Kriener, K. Segawa, Z. Ren, S. Sasaki, and Y. Ando, 
Phys. Rev. Lett. {\bf 106}, 127004 (2011). 
\bibitem {PhysRevLett.105.097001}%
  L.~Fu\ and\ E.~Berg,\ Phys. Rev. Lett.\ \textbf {105},\ 097001 (2010).
\bibitem {PhysRevLett.107.217001}%
  S.~Sasaki, M.~Kriener, K.~Segawa, K.~Yada, Y.~Tanaka, M.~Sato, \ and\
  Y.~Ando,\ Phys. Rev. Lett.\ \textbf {107},\ 217001 (2011).
\bibitem {PhysRevB.86.064517}%
  T.~Kirzhner, E.~Lahoud, K.~B.\ Chaska, Z.~Salman, \ and\ A.~Kanigel,\ Phys.
  Rev. B\ \textbf {86},\ 064517 (2012).
 \bibitem{PhysRevLett.110.117001}
 N. Levy, T. Zhang, J. Ha, F. Sharifi, A. A. Talin, Y. Kuk, and J. A. Stroscio, 
 Phys. Rev. Lett. {\bf 110}, 117001 (2013). 
 \bibitem{Matano}
 K. Matano, H. Yamamoto, K. Ueshima, F. Iwase, M. Kriener, K. Segawa, Y. Ando, G.-q. Zheng, 
 JPS 2014 Annual (69th) Meeting, 28aBF-12. 
  %
 \bibitem{Hashimoto}
 T. Hashimoto, K. Yada, A. Yamakage, M. Sato and Y. Tanaka, J. Phys. Soc. Jpn. {bf 82}, 044704 (2012). 
 \bibitem{NagaiMajo}
 Y. Nagai, H. Nakamura, and M. Machida, arXiv:1211.0125v2 to be published in J. Phys. Soc. Jpn. 
 \bibitem{NagaiThermal}
 Y. Nagai, H. Nakamura, and M. Machida, Phys. Rev. B {\bf 86}, 094507 (2012).  
\bibitem{Hayashi}
N. Hayashi, M. Ichioka and K. Machida, Phys. Rev. B {\bf 56}, 9052 (1997). 
 \bibitem{NagaiJPSJ}
Y. Nagai, Y. Ueno, Y. Kato and N. Hayashi, J. Phys. Soc. Jpn. {\bf 75}, 104701 (2006).
\bibitem{NagaiJPSJCe}
Y. Nagai, Y. Kato and N. Hayashi, J. Phys. Soc. Jpn. {\bf 75}, 043706 (2006). 

\bibitem{Hess}
H. F. Hess, R. B. Robinson and J. V. Waszczak, Phys. Rev. Lett. {\bf 64}, 2711 (1990). 
\bibitem{Nishimori}
H. Nishimori, K. Uchiyama, S. Kaneko, A. Tokura, H. Takeya, K. Hirata and N. Nishida, J. Phys. Soc. Jpn. {\bf 73} 3247 (2004). 
\bibitem{Gygi}
F. Gygi and M. Schluter, Phys. Rev. B {\bf 43}, 7609 (1991). 
 \bibitem{Sakakibara}
 T. Sakakibara, A. Yamada, J. Custers, K. Yano, T. Tayama, H. Aoki and K. Machida, 
 J. Phys. Soc. Jpn. {\bf 76}, 051004 (2007). 
 \bibitem{Matsuda}
 Y. Matsuda, K. Izawa, and I. Vekhter, J. Phys. Condens. Matter {\bf 18}, R705 (2006). 



\bibitem{NagaiQuasi}
Y. Nagai, H. Nakamura, and M. Machida, J. Phys. Soc. Jpn. {\bf 83}, 053705 (2014). 
\bibitem{NagaiPRL}
Y. Nagai, and N. Hayashi, Phys. Rev. Lett. {\bf 101}, 097001 (2008). 
\bibitem{NagaiYNi}
Y. Nagai, Y. Kato, N. Hayashi, K. Yamauchi, and H. Harima, Phys. Rev. B {\bf 76}, 214514 (2007). 
\bibitem{Hao}
L. Hao and T. K. Lee, Phys. Rev. B {\bf 83}, 134516 (2011).
\bibitem{Yamakage}
A. Yamakage, K. Yada, M. Sato, and Y. Tanaka, Phys. Rev. B {\bf 85}, 180509 (2012). 
\bibitem{Mizushima}
T. Mizushima, A. Yamakage, M. Sato, and Y. Tanaka, arXiv:1311.2768 (unpublished). 

\bibitem{KatoJPSJ}
Y. Kato, J. Phys. Soc. Jpn. {\bf 69}, 3378 (2000).
\bibitem{Nagato}
Y. Nagato, K. Nagai, and J. hara, J. Low Temp. Phys. {\bf 93}, 33 (1993).
\bibitem{Higashitani}
S. Higashitani and K. Nagai, J. Phys. Soc. Jpn. {\bf 64}, 549 (1995). 
\bibitem{NagatoHigashitani}
Y. Nagato, S. Higashitani, K. Yamada, and K. Nagai, J. Low Temp. Phys. {\bf 103}, 1 (1996). 
\bibitem{SchopohlMaki}
N. Schopohl and K. Maki, Phys. Rev. B {\bf 52}, 490 (1995). 
\bibitem{Schopohl}
N. Schopohl, arXiv:cond-mat/9804064 (unpublished). 
\bibitem{Melnikov}
A. S. Melnikov, D. A. Ryzhov, and M. A. Silaev, Phys. Rev. B { \bf 78}, 064513 (2008). 
\bibitem{NagaiOrganic}
Y. Nagai, H. Nakamura, and M. Machida, Phys. Rev. B {\bf 83}, 104523 (2011). 
\bibitem{HayashiPRL}
N Hayashi, M Ichioka, and K Machida, Phys. Rev. Lett. {\bf 77}, 4074 (1996). 
\end{thebibliography}
\end{document}